\documentclass[aps,twocolumn,nofootinbib,superscriptaddress,showpacs]{revtex4}
\usepackage{graphicx}
\usepackage{amsfonts}
\usepackage{psfrag}
\usepackage{amsmath}
\usepackage{float}
\usepackage{amssymb}  

\def\be{\begin{equation}}
\def\ee{\end{equation}}
\def\ba{\begin{eqnarray}}
\def\ea{\end{eqnarray}}

\def\Tr{\text{Tr}}

\def\ut#1{\rlap{\lower1ex\hbox{$\sim$}}#1{}}

\newcommand{\R}{\mathbb{R}}
\DeclareFontFamily{U}{rsfs}{}         
\DeclareFontShape{U}{rsfs}{m}{n}{<5> rsfs5 <6><7> rsfs7          %
  <8><9><10><10.95><12><14.4><17.28><20.74><24.88> rsfs10}{}     %
\DeclareMathAlphabet{\mathfs}{U}{rsfs}{m}{n}                     %
\newcommand{\mfs}[1]{\mathfs {#1}}                               %
\newcommand{\n}{{\nonumber}}

\newcommand{\sH}{{\mfs H}}

\newcommand{\sN}{{\mfs N}}

\newcommand{\sI}{{\mfs I}}

\begin{document}

\title{Statistical and entanglement entropy for black holes in quantum geometry }

\author{Alejandro Perez}
\affiliation{
    Aix Marseille Université, CNRS, CPT, UMR 7332, 13288 Marseille, and
    Université de Toulon, CNRS, CPT, UMR 7332, 83957 La Garde, France.
}

\begin{abstract}
We analyse the relationship between entanglement (or geometric) entropy with statistical mechanical entropy of horizon degrees of freedom when described in the framework of isolated horizons in loop quantum gravity. We show that, once the relevant degrees of freedom are identified, the two notions coincide. The key ingredient linking the two notions  is the structure of quantum geometry at Planck scale implied by loop quantum gravity, where correlations between the {\em inside} and {\em outside} of the black hole are mediated by eigenstates of the horizon area operator.   
\end{abstract}


\maketitle

In the semiclassical regime where an approximate notion of black hole (BH) makes sense, horizon area  is quantized in loop quantum gravity (LQG) \cite{Perez:2004hj, lqg1,lqg2}. As a consequence, the dimensionality of the surface Hilbert space, compatible with a maximum macroscopic area, is finite and grows exponentially with the area in the large area limit \cite{Rovelli:1996dv, ABK}. In other words, if one treats the horizon itself as a thermodynamical system in equilibrium---justified in the case of large semiclassical black holes---then the micro canonical entropy, in the ensemble of surface states, is proportional to the macroscopic black hole area. The relevant ensemble is the one defined by the horizon surface states.

The standard loop quantum gravity counting yields \cite{Meissner:2004ju, Agullo:2008yv}
\be\label{uni}
S_{stat}=\frac{\gamma_0}{\gamma \frac{G}{G_N}}\frac{A}{4G_N\hbar}
\ee   
where $G$ is the UV value of the gravitational constant at the fundamental scale, $G_N$ is the low energy Newton's constant, $\gamma$ is the Immirzi parameter,
and $\gamma_0$ is a numerical constant that appears in the asymptotic expression of the number of states. 

The above result is compatible with Bekenstein-Hawking's entropy only if one fixes \be\label{gaga} \gamma=\gamma_0 \frac{G_N}{G}.\ee
The necessity of fixing $\gamma$  in order to achieve consistency with the semiclassical regime is puzzling.  The reason is that the Immirzi  parameter is a  topological coupling in the gravity first order action  not affecting the classical equations of motion (like  the $\theta$ parameter of Yang-Mills theories \cite{Date:2008rb, Rezende:2009sv}).  Therefore, in contrast with expression (\ref{uni}), one would expect to find $\gamma G$ only in quantum corrections of a leading term matching exactly the Bekenstein-Hawking entropy (as the Immirzi parameter plays an important role in the quantum theory where geometric operators have discrete spectra in the fundamental area $\ell^2_{LQG}=\gamma G\hbar$).

One logical possibility \cite{Jacobson:2007uj} is to interpret equation (\ref{gaga}) as a renormalization condition fixing the relationship between $G_N$ and $G$. If that were  the case, then the numerical value $\gamma_0$ would have to have a clear interpretation in the renormalization
group flow of gravity. However, due to the completely combinatorial way in which $\gamma_0$ arises (which does not make reference to any dynamical notion) it is so far unclear how such scenario could be realized. 

There is however a recent new perspective that might be helpful in resolving this question.
The idea is that vacuum fluctuation in the non geometric sector (including matter degrees of freedom) should be 
appropriately taken into account in the computation of black hole entropy (as we will argue in Section \ref{aqua} all degrees of freedom close to the horizon must be accounted for). This is not so in the treatment leading to (\ref{ent}) as only pure-geometry area eigenvalues are counted while ignoring the degeneracy associated to other degrees of freedom (including matter). Indeed by taking these degrees of freedom suitably into account one obtains\cite{Ghosh:2013iwa}, for arbitrary values of $\gamma$, 
\be\label{estata}
S_{stat}=\frac{A}{4G_N\hbar}+ \eta \frac{\sqrt{A}}{\sqrt{\gamma G\hbar}}
\ee
where $\eta$ a dimensionless constant that depends on the punctures statistics.
As expected the fundamental scale $\gamma G \hbar$ appears only in the quantum corrections to the Bekenstein-Hawking entropy.

The fundamental surface degrees of freedom associated to matter fields are hard to describe in LQG. Thus vacuum fluctuations in the analysis leading to the previous equation are accounted for  by 
using (qualitative) information provided by standard QFT on the black hole background (in a spirit analogous to what is done in cosmological situations \cite{Agullo:2012fc}).
In particular there is an expected contribution to black hole entropy from the entanglement entropy of quantum fields across the black hole horizon \cite{Bombelli:1986rw}.
In standard QFTs the result is 
\be\label{ent}
S_{ent}=\lambda {\frac{ A}{G_N\hbar}}+corrections.
\ee
with $\lambda$ left undetermined 
due to UV divergences and other ambiguities such as the species problem (concretely, $\lambda =\lambda_0 G_N \hbar \epsilon^{-2}$ for $\lambda_0$ a regularization dependent dimensionless constant and $\epsilon$ a UV cut-off length).

In the analysis of \cite{Ghosh:2013iwa} equation (\ref{estata}) follows from the assumption that the degeneracy contribution to the area spectrum coming from vacuum fluctuations of non geometric degrees of freedom is given by the exponential of (\ref{ent}) with an undetermined $\lambda$ (this justifies the term {\em qualitative} used above 
\footnote{This type of holographic degeneracy appears naturally when considering the analytic continuation of the pure geometric degeneracy  from real the Immirzi parameter $\gamma$ to $\gamma\to \pm i$ \cite{Frodden:2012dq, Han:2014xna}. The same holographic behaviour of the number of degrees of freedom available at the horizon surface
is found from a conformal field theoretical perspective for $\gamma=\pm i$ \cite{Ghosh:2014rra}. A relationship between the termal nature of BH horizons and self dual variables seems also valid according to similar analytic continuation arguments \cite{Pranzetti:2013lma}.}).  In the perturbative quantum gravity framework there are indications that  the ambiguities encoded in the value of $\lambda$ disappear if
the gravitational degrees of freedom are correctly taken into account \cite{Bianchi:2012br}. In 
the non perturbative framework, 
such possibility is reinforced in a completely independent way by the results of \cite{Ghosh:2013iwa} where consistency fixes $\lambda=1/4$ up to quantum corrections. 

However, as just mentioned, in the argument that leads to (\ref{estata}) one assumes that one can interpret the (exponential of the) entanglement entropy of quantum fields across the horizon as a measure of the degeneracy of the area spectrum of the horizon (including all the UV degrees of freedom). This assumption necessitates a suitable relationship between entanglement and statistical entropy. Exploring and establishing the degree to which such relationship holds in the LQG treatment of BHs is one of the aims of this work. 
Some aspects concerning this question have been considered in previous literature (for a recent discussion see \cite{Bodendorfer:2014fua}). Entanglement of apparent horizons in toy theories have been considered in \cite{Husain:1998zw},  regions of spin network states have been studied in \cite{Donnelly:2008vx}, and it is an important part of the argument of \cite{Pranzetti:2013lma}. 

Finally, we would like to point out in this introduction that there is a well know relationship between entanglement entropy  and Wald's entropy \cite{Wald:1993nt} (for a review see \cite{Solodukhin:2011gn}).  
The present analysis deals with a similar question but at the more fundamental Planckian level where a statistical mechanical account of the thermodynamical properties of BHs becomes available (including the fully quantum treatment of  gravitational contributions). 

In Section \ref{next} we will argue that the relevant degrees of freedom both for entanglement and statistical mechanical approaches to BH entropy are to be found on the local Planckian vicinity of the BH horizon. In making this case, we will revisit the results of \cite{Bianchi:2012br}, and gain new insights by discussing their generalisation to eternal static black holes in Section \ref{bhbh}. In Section \ref{IH} we will briefly review the quantum description of the BH horizon quantum degrees of freedom in LQG. With this framework at hand we will establish the sense in which entanglement and statistical entropy are equivalent in quantum geometry in Section \ref{sf}.

\section{why surface degrees of freedom?}\label{next}

In this section we will discuss a common feature between 
the entanglement approach to black hole entropy and the statistical mechanical approach followed in loop quantum gravity:
 that the relevant correlations are the UV correlations across the BH horizon in the first case; that the relevant microstates are local excitation of the horizon geometry and matter on the horizon in the second case.  
In both cases a separation of scales is necessary to isolate the relevant  
physics behind the notion of BH entropy. Once this ingredient is incorporated to the analysis
we will show that both approaches actually compute the same thing: namely, the Boltzmann-Gibbs 
entropy of the ensemble of surface microstates. Therefore, from the appropriate perspective, BHs are standard thermodynamical objects where the horizon plays the role of the statistical mechanical system. 

\subsection{Long range correlation in entanglement entropy}\label{bhbh}

Let us start with some discussion of the entanglement approach.
In particular we want to point out the influence of global features of the 
quantum state considered when trying to compute the entanglement entropy associated to the presence of a horizon.

The Rindler wedge with its family of uniformly accelerated observers
and associated Rindler horizon captures some of the physical aspects
of black hole systems in their infinite area limit. Entanglement entropy of the Minkowski vacuum 
for the wedge is ill defined for the UV problems mentioned above or  
exactly zero if one uses the regularization prescription that is consistent with the semiclassical Einstein's equations
compatible with the flat background (i.e. for which $\langle T_{\mu\nu}\rangle=0$ \cite{Wald:1995yp}). We will come back to this important point in Section \ref{aqua}. However, the changes in the 
entanglement entropy (relative entropy \cite{Casini:2008cr}) under perturbations of the reduced density matrix $\rho=\rho_0+\delta\rho$ turn out to be well defined \cite{Bianchi:2012br} and give \ba\label{dent}
\delta S_{ent}&=&-{\Tr}[(\rho_0+\delta\rho)\log(\rho+\delta\rho)]\n \\&=&
2\pi \int_{\Sigma} \delta \langle T_{\mu\nu}\rangle \chi^\mu d\Sigma^{\nu}
\n \\&=&{\frac{ \delta A}{4G_N\hbar}}
\ea
where $\chi$ is the corresponding boost killing vector field (to which accelerated observers are tangent), $\Sigma$ is an arbitrary Cauchy surface of the Rindler wedge, and $\delta A$ is the change in the Rindler horizon area produced by the back reaction of the perturbation.
The second line follows from the fact that, formally, the reduced density matrix $\rho_0$ obtained from the Minkowski pure state by tracing out the degrees of freedom inaccesible to the Rindler accelerated observers takes the form
\be\label{roo}
\rho_0=\frac{\exp (- 2\pi \int_{\Sigma} \hat T_{\mu\nu} \chi^\mu d\Sigma^{\nu})}{Z}.
\ee
Conservation of the current $\delta  \langle T_{\mu\nu}\rangle \chi^\mu$  can be used to express the second line as the energy momentum flux of the perturbation across the horizon. The Raychaudhuri equations for the
horizon generators---which are a consequence of semiclassical Einsteins equations---imply the final result. 
In this sense the above equation is semiclassical and hence insensitive to the UV quantum gravity physics (as expected for a relative entanglement entropy).

One might have naively expected an additive term $\delta S_{\infty}$ coming from the energy momentum flux to infinity in the direction normal to the horizon. However,  as this corresponds to (a piece of) a single generator of $\sI^{+}$ (future null infinity)  the associated energy flux vanishes for a regular perturbation (e.g. compact support on $\Sigma$). A proof of such statement for massless scalar fields is given in \cite{Wald:1995yp}, section 5.1. On physical grounds one expects this to be valid in general. However, we will see below that this potential entropy flux at infinity becomes non trivial when equation (\ref{dent}) is generalised to the black hole setting.  

\begin{figure}[h] \centerline{\hspace{0.5cm} \(
\begin{array}{c}
\psfrag{mytag}{$i_0$}
\psfrag{b}{I}\psfrag{bb}{\bf I}\psfrag{a}{\bf II}\psfrag{aa}{\bf II}
\psfrag{q}{$q$}\psfrag{p}{$q^{\prime}$}\psfrag{d}{$\sI^{+}$}\psfrag{e}{$\sI^{-}$}
\psfrag{c}{$i_0$}\psfrag{g}{$i^{+}$}\psfrag{f}{$i^{-}$}
\includegraphics[width=4cm]{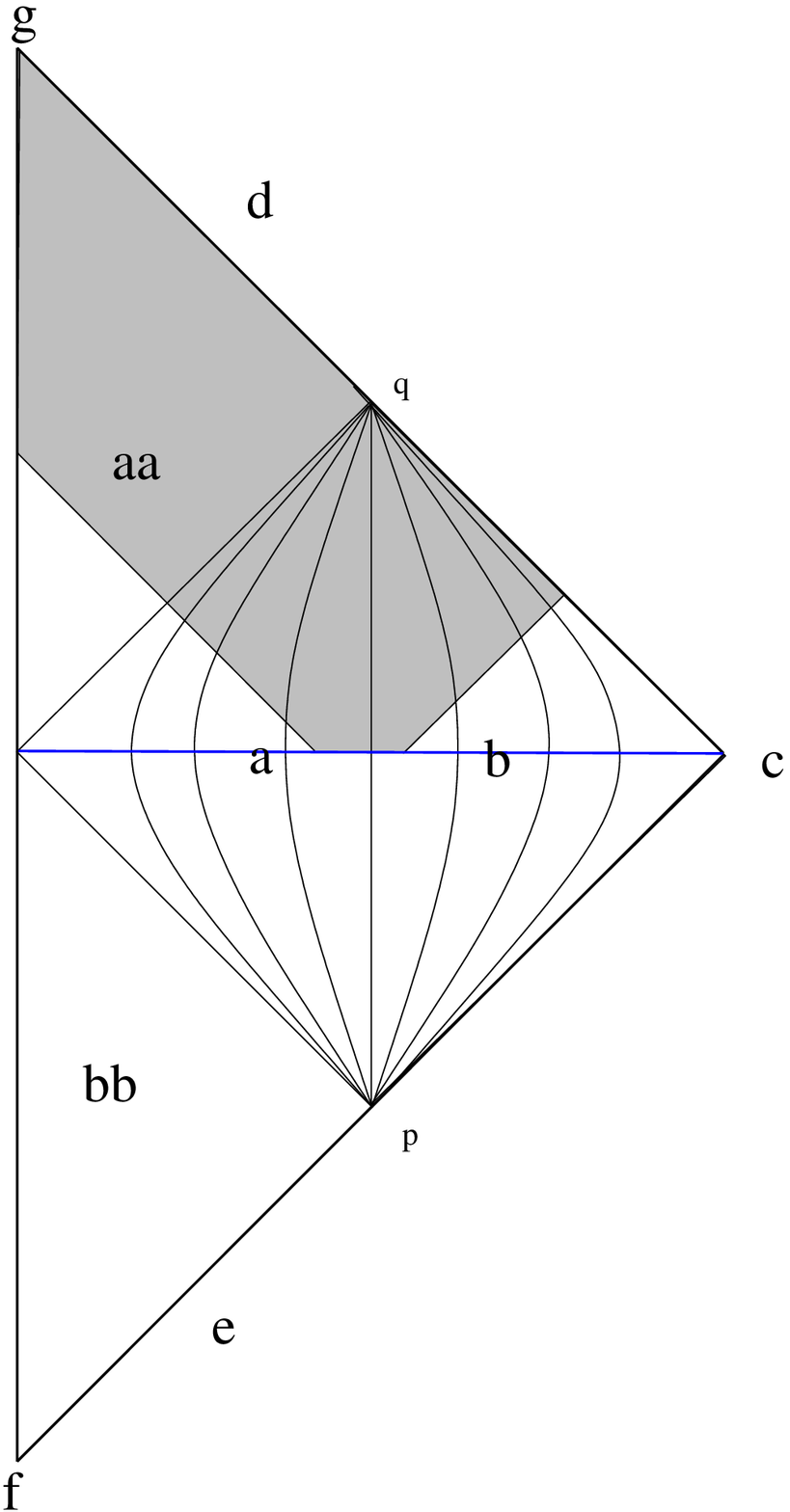}
\end{array}\ \ \ \ \ \ \ \ \ \ \ \ \ \ \ \ \ \ \ \ \ \ \ 
\begin{array}{c}
\psfrag{b}{I}\psfrag{bb}{\bf I}\psfrag{a}{\bf II}\psfrag{z}{\bf }
\psfrag{q}{$q$}\psfrag{p}{$q^{\prime}$}\psfrag{d}{$\sI^{+}$}\psfrag{e}{$\sI^{-}$}
\psfrag{c}{$i_0$}\psfrag{g}{$i^{+}$}\psfrag{f}{$i^{-}$}
\psfrag{b}{\bf I}\psfrag{bb}{\bf I}\psfrag{a}{II}\psfrag{aa}{\bf II}
\psfrag{da}{$\sI^{\prime+}$}\psfrag{ea}{$\sI^{\prime-}$}
\psfrag{ca}{$i'_0$}\psfrag{ga}{$i^{\prime +}$}\psfrag{fa}{$i^{\prime-}$}
\includegraphics[width=7.5cm]{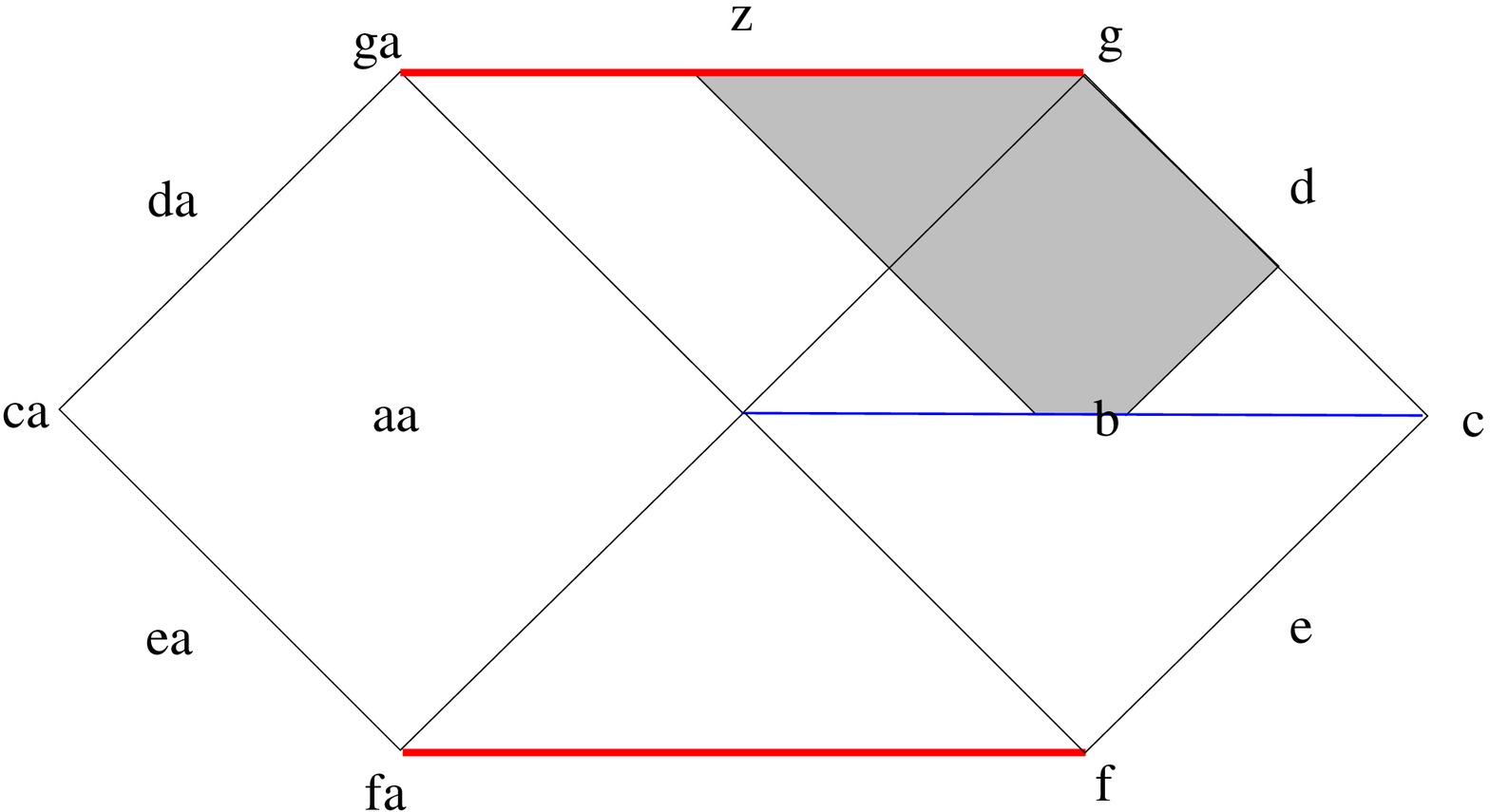}
\end{array}
\)}
\caption{{\bf On the left:} Penrose diagram of Minkowski spacetime exhibiting a Rindler horizon. In the conformal compactification the Rindler horizon is the past light cone of a point $q\in \sI^{+}$ and the future light cone of a point $q^{\prime}\in \sI^{-}$ minus a single generator of $\sI^{+}$ and $\sI^{-}$ respectively. This global picture implies that, for suitable fields and in contrast with a BH horizon, the Rindler horizon is a good initial value surface. The 
Rindler wedge corresponds here to region I. In the Rindler wedge portion of the diagram, bold face letters {\bf II} in contrast to simply I tell us that I is separated from the readers `view' by the Rindler horizon. {\bf On the right:} The Penrose diagram of the eternal Schwarzschild BH, where radiation can escape to infinity without being registered on the BH horizon. The shaded regions represent a perturbation that has compact support at a Cauchy surface of region {\bf I}: on the left, the energy flow of the perturbation cannot avoid crossing the Rindler horizon.   
} \label{rindler}
\end{figure}

While it is true that a finite slab of proper width $\ell$ outside a stationary BH horizon is locally isometric with the corresponding slab of width $\ell$ of Rindler space-time in the limit where the BH area $A\to \infty$ while keeping  $\ell$ fixed \cite{Frodden:2011eb}; the two space times are clearly not the same globally in that limit. The difference is apparent when one considers the spacetime conformal compactification as in Figure \ref{rindler}. This is a key limitation of modelling black hole physics with a Rindler spacetime in  quantum field theory where long range correlations can be important.

As mentioned, there is no entanglement entropy flow term at infinity in the Rindler analysis: 
due to the global structure of the Rindler horizon regular perturbations cannot avoid crossing the horizon and register their entropy content by an area change as in (\ref{dent}).
However, the situation completely changes in the context of BHs. Notice first that the argument leading to (\ref{dent}) can be generalized in order to calculate perturbations to the entanglement entropy of the eternal static black hole in the Hartle-Hawking vacuum. This is achieved  
by replacing (\ref{roo}) with $\rho_0=\exp(-2\pi\kappa^{-1} \int_{\Sigma} \hat T_{\mu\nu} \chi^\mu d\Sigma^{\nu})$  where $\chi^u$ is the Killing field generating inertial time translations at infinity (such simple generalization might not be available for general stationary black holes where the analog of the Hartle-Hawking vacuum does not exist \cite{Kay:1988mu}). However, the BH generalization of (\ref{dent}) will contain a non vanishing extra term accounting for the energy-momentum 
flow of the perturbation at infinity, namely       
\ba\label{dentada}
\delta S^{BH}_{ent}={\frac{ \delta A}{4G_N\hbar}}
+\delta S_{\infty},
\ea
where $\delta S_{\infty}$ is the entropy flow across future null and timelike infinity  $\sI^{+}\cup i^+$.

 \onecolumngrid
 
\twocolumngrid

Thus equation (\ref{dentada}) tells us two important pieces of information. The first one is that long range correlations contribute non trivially to entanglement entropy but that these contributions are not related to BH entropy.  Notice that one can have $\delta S_{ent}\not=0$ with a perturbation that does not affect the horizon surface (the entanglement boundary) by arranging $\delta \langle T_{\mu\nu}\rangle$ to completely flow into $\sI^{+}$ without crossing the horizon.  Long range correlations---see equation (\ref{had})---which are present in massless fields can produce entropy flow to null infinity as well as to future timelike infinity
(the classical counterpart of these are the so-called tails). Massive fields cannot produce entropy flow out to $\sI^{+}$; however, for suitable perturbations (unbounded states of the perturbation) there are non trivial contributions to $\delta S_{\infty}$ from the flow through $i^+$. The term at infinity in (\ref{dentada}) can be written as%
\footnote{\label{foot}More generally, the flow of entropy across $\sI^+$ can also be given a thermodynamical interpretation for stationary local observers inside as follows: assume that one gives initial conditions for the perturbation $\delta\rho$ on a Cauchy surface $\Sigma$ for the exterior of the static BH space-time. Assume that  $\langle\delta T_{\mu\nu}\rangle={\rm Tr}[\hat T_{\mu\nu}\delta \rho]$ has compact support on $\Sigma$. Take a timelike hyper surface $W$ defined by $r=r_0=$constant intersecting $\Sigma$ on a sphere with $r_0$ sufficiently large to contain the region  where $\langle\delta T_{\mu\nu}\rangle\not=0$. $W$ represents a  family of stationary observers surrounding the BH. Then one has that
$\delta S^{\sI^+}_{\infty}=-{\delta E}/{T}$ 
where $\delta E$ is the energy-momentum flow across $W$, and
$T=T_{H}/\|\xi\| $ is the local temperature of the Hartle-Hawking state as measured by the stationary observers at $W$ (with $\|\xi\|$ the norm of the stationarity Killing field normalised at infinity). This is the the standard thermodynamical entropy flow of the perturbation across $W$ as measured by these local observers.
}
\be
\delta S_{\infty}=\frac{\delta E_{+}}{T_{H}},
\ee 
where $\delta E_+$ is the amount of energy flow to future infinity and $T_{H}=\kappa/(2\pi)$ is Hawking temperature.
One can also write (\ref{dentada}) as
\be\label{huno}
\delta S^{BH}_{ent}=\frac{\delta M}{T_H}={\frac{ \delta A}{4G_N\hbar}}
+\frac{\delta E_{+}}{T_{H}},
\ee 
where $\delta M$ is the change in the ADM mass of the system due to the perturbation.
Thus, even thought it is well known that bulk corrections exist in (\ref{ent}), the present analysis makes their geometric origin and thermodynamical meaning transparent.

All this implies that if one wants to describe black hole entropy from the entanglement idea one has to have the means of separating the bulk contributions to entanglement from the genuine horizon contributions. The criterion is locality: the term arising from the UV physics close to the entanglement boundary are the ones to be taken into account. Concretely, in standard QFT these are the modes that produce a UV diverging contribution or the ones that lead to a term proportional to the area. As we will discuss further below, the same separation of degrees of freedom is made in the statistical mechanical account of BH entropy in LQG.

The second indication from the perturbative result (\ref{dent})---the main message of \cite{Bianchi:2012br}---is that the inclusion of the gravitational degrees of freedom 
renders entanglement entropy finite and free from its usual regularization ambiguities and the species problem.
This second conclusion is validated  in the non perturbative by the analysis of \cite{Ghosh:2013iwa} (given the relationship between entanglement and the number of surface microstates advocated here). The argument  shows  how the equations of state of the quantum horizon imply that the undetermined (UV divergent) constant $\lambda$ (of equation (\ref{ent})) is finite an takes the value $\lambda=1/4$ (up to quantum corrections) in the semiclassical regime.  As suggested in \cite{Bianchi:2012br}, gravitational effects resolve the ambiguity of the standard QFT entanglement account. However, as mentioned before, one key assumption in \cite{Ghosh:2013iwa} is  that the equation (\ref{ent}) can be directly interpreted
as degeneracy of the area spectrum of the quantum horizon $D(A)=\exp(\lambda {A}/{(G_N\hbar)})$ due to non-geometric degrees of freedom. 
Before arguing that this is indeed correct let us discuss the question of the localisation of the relevant degrees of freedom accounting for BH entropy.

\subsection{From the statistical mechanics perspective the relevant degrees of freedom are localised at the horizon}\label{aqua}

Mixed states do not exist in nature. However, the concept is the key mathematical tool for the description of systems
in statistical mechanics which in turn is the fundamental basis of thermodynamics. One of its keystones in the construction is the introduction of a notion of coarse graining producing exact physical laws 
for macroscopic variables by treating microscopic details probabilistically. Statistical mechanics becomes useful in those situations where one can concentrate on suitable average quantities assumed to take well defined values up to (typically) small fluctuations whose exact details are ignored.  The states of such systems are mathematically modelled by a statistical mixture of microstates (in a suitably defined ensemble)  in quantum statistical mechanics. It is in terms of the latter that the notion of 
thermodynamic entropy becomes non trivial and meaningful. 

Stationary black holes are, in a suitable sense, equilibrium statistical mechanical systems. As stationary space times they represent the final stage of gravitational collapse which is completely described, according to the no-hair theorem,  by
members of the Kerr-Newman family labelled by only three macroscopic parameters: 
the mass $M$, the angular momentum $J$, and the electromagnetic charge $Q$.
These three macroscopic parameters are the coarse graining variables for the thermodynamical description of BHs.
 
The microstates are to be found in the local Planckian details of fields at the BH horizon. Stationary observers {\em outside} of the BH have causal access to the local degrees of freedom of matter and space-time as they crossed the forming BH horizon. These  degrees of freedom are entirely registered on the horizon; as a null surface whose domain of dependence is the interior of the BH. However, they are (infinitely) redshifted to late stationary {\em outside} observers  who  are, in this way,  only sensitive to their UV or Planckian structure.   The  many different initial conditions leading to the formation of a black with the same $M$, $J$ and $Q$ are ironed out by this infinite redshift and become inaccessible in practice to macroscopic observers (much like the phase space position of molecules in a gas in thermal equilibrium). Moreover, because of the redshift factor vanishing at the horizon, these excitations appear as horizon {\em surface degrees of freedom} for (late) stationary {\em outside} observers. 
The latter include both the Planckian geometric as well as matter  excitations (`vacuum fluctuations').
These are the micro states in the statistical mechanical description of BHs; they are, by the above argument, fundamentally quantum in nature, and they are localised at the BH horizon.  

The thermal system is the horizon itself. This view is complemented by the fact that Hawking's effect of BH radiation is entirely due to the near horizon geometry in stationary BHs (for a concrete example see \cite{huang}). 
 
\subsubsection{A perspective on information loss in BH systems} \label{info}

According to the perspective advocated here (aiming at the statistical mechanical description of BH thermodynamics), a system producing a BH by gravitational collapse is to be described by a mixed state initially. The coarse graining variables are the final parameters $M, J$, and $Q$ of a stationary BH,
\footnote{As, in the context of the information loss paradox discussion, the system is analyzed until complete evaporation of the BH to Hawking radiation; here $M, J$, and $Q$ are the values of these parameters long time after the collapse but before back reaction of the hawking radiation starts being important.}
while the large number of different initial conditions leading to it at late times are the microstates.
As argued the latter are Planckian in nature and involve both matter as well as gravitational degrees of freedom.

In this context, dynamical evolution from the initially mixed-state need not to purify it in the late future. 
Even if the standard semiclassical expectation is correct and one ends up with a final state made purely of the thermal radiation emitted during BH evaporation, this will not represent by itself a paradox. 
In fact if the evolution is unitary
one would expect the initial entropy of the mixed state on $\sI_-$  simply to match that of the final state at $\sI^{+}$. 

However, this is not yet satisfactory to clarify the fate of information `falling behind the BH event horizon'. This is so because one can set up a thought scenario where the initial state is indeed pure (one of the microstates leading to the formation of the macroscopic BH in late times). However, as mentioned above, this would necessitate the specification of Planckian degrees of freedom; therefore, the use of a  quantum dynamical evolution that can only be described in a fully background independent manner (semiclassical arguments are just not the right language to describe the question of unitarity). If evolution is unitary
\footnote{The possibility that information is lost to outside observers by quantum gravity in the context of gravitational collapse cannot be ruled out and does not represent any inconsistency with local unitarity \cite{Wald:1999vt}. In order to  a quantum gravity dynamical description of the fate of the classical BH singularity is mandatory.}
 the end result would look---from the semiclassical perspective---as a thermal state for continuum field excitations (particles of QFT) but it would remain pure due to the correlations with the Planckian gravitational degrees of freedom that remain hidden to low energy (coarse) observers.  If evolution is unitary
 a detailed account on how the correlations leak out of `the BH region' during the long history of BH formation and subsequent evaporation would have to be given in the framework of the quantum theory. However, we have seen in the discussion of the present paper that non-local correlations are relevant in the discussion of entanglement between the {\em inside} and the {\em outside} of the horizon. It seems to us that here is no need to invoke other sources of non-locality such as the constraint structure of gravity as advocated  in a proposal aiming at preserving unitarity in the much stronger sense of the ADS-CFT framework see \cite{JACOBSON:2013ewa}.

The scenario described here seems quite plausible in the framework LQG, where even flat space time is expected to arise as very intricate linear combinations of polymer like quantum excitations. In such framework, 
the very notion of  Minkowski  is only approximate, i.e., arising in a suitable coarse grained sense \cite{Varadarajan:1999it, Ashtekar:2001xp, Ashtekar:2002sn}. Thus many different microstates will be classified as flat spacetime by low energy (standard QFT) observers for which Planckian details and correlations with the fundamental degrees of freedom remain hidden.   

The picture proposed here is quite the analog of that of burning an encyclopaedia: information before and after is the same due to unitarity.  Before burning we have the books in the atmosphere in some initial state. After burning the information remains in the correlations of the individual molecules of hot gas produced by the combustion.  If we wait even longer, then the air in the atmosphere seems to recover its initial state which of course is not the same due to the everlasting correlations of the microscopic degrees of freedom. Observers that are not capable of reading these correlations must accept with resignation that information is (practically) lost. 
  
%

\subsection{Black hole entropy as entanglement entropy: the relevant correlations are also the local ones across the horizon}

%
In quantum field theory vacuum correlations at space-like separated points in $3+1$ dimensions diverge quadratically in the 
geodesic distance $d(x,y)$ separating them, namely 
\be\label{had}
\langle\phi(\vec x)\phi(\vec y)\rangle\approx \frac{1}{|d(x,y)|^2}.
\ee
As a consequence of the local nature of correlations  entanglement entropy 
in $3+1$ dimension (when calculated in standard QFT) diverges quadratically with 
the cut-off scale $\epsilon$ and is hence proportional to the boundary surface area $A$:
\be
S_{ent}=\lambda_0 \frac{A}{\epsilon^2}+corrections,
\ee 
where $\lambda_0$ is a regularisation dependent factor. The previous equation is  the main motivation to view entanglement entropy as a possible candidate for an account BH entropy \cite{Bombelli:1986rw}.  It can be proved rigorously for ground states of lattice systems with suitable interactions mimicking QFT systems \cite{Eisert:2008ur}.
Despite of the fact that entanglement entropy is not really well defined in the standard QFT scenario, 
entropy difference between two quantum states (relative entropy) is well defined and satisfies important properties \cite{Bousso:2014sda, Casini:2008cr}. 
However, a description of entanglement entropy across the horizon in view of a fundamental explanation of BH entropy cannot be based on the concept of relative entropy in a background independent approach as the one of LQG. This is thus because relative entropy is insensitive to the UV physics that we have argued encodes the relevant degrees of freedom behind BH entropy. Thus, for the problem at hand, one necessitates a full quantum gravity formulation, or UV completion, that eliminates the ambiguities of entanglement entropy referred to above
\footnote{One would expect that gravitational effects would modify correlations at the microscopic scales.
Interestingly if one bounds the correlations (\ref{had}) by $\ell^{-2}_{Pl}$ then the Hawking effect is not affected 
in the large BH limit \cite{Agullo:2009wt}. Such bound for correlations are confirmed by spherically symmetric model calculations in LQG \cite{Gambini:2013nea} \footnote{It is interesting to consider that such bound on Lorentz invariant correlations is perhaps the correct language approach the problem of Lorentz invariance versus discreteness in quantum gravity \cite{Collins:2004bp}. The UV problem of standard local quantum field theory hints to the idea of a completion of the microscopic  physics by quantum gravity were the basic discrete elements should be null.}. Here we are assuming that 
the fundamental discreteness predicted by LQG must be compatible with Lorentz invariance as enforced on physical grounds by the constraints raised in \cite{Collins:2004bp} and rediscussed from an independent perspective in \cite{Polchinski:2011za}. } 
.  In what follows we describe what quantum geometry and LQG tell us about entanglement from the UV structure close to the BH horizon.

\section{Accounting for surface degrees of freedom in loop quantum gravity}\label{IH}

In this section we describe the isolated horizon model for the black hole horizon 
surface degrees of freedom. We will also show that the entanglement entropy 
of these degrees of freedom coincides in a suitable sense with the micro canonical entropy 
of the ensemble defined by the surface micro states. 

\subsection*{Statistical mechanics of the Horizon degrees of freedom}\label{lqg}

Isolated horizons \cite{AKLR} is the standard mathematical framework for the statistical mechanical treatment of quantum black holes in LQG. The notion is an idealisation that allows to represent a suitable quasi-local definition of BH horizons in equilibrium by an infinite dimensional phase space to be quantized. While such framework might not capture all the physics associated to BHs in quantum gravity (in particular due to its local nature it is unclear how to encode in the formalism the long range correlations that are important in the description of certain quantum processes),  it is expected to describe the local microscopic degrees of freedom necessary for the statistical mechanical account of semiclassical black holes. 

The presentation of the basic geometric idea will be enough for the purposes of this letter. The basic idea behind isolated horizons can be  understood from the perspective of  the initial value problem of general relativity in terms of characteristic data on null surfaces.
One considers a null surface $\Delta$ with topology $S^2\times [0,1]\subset \R$ and a transversal null surface $\delta$ as shown in Figure \ref{two}. For simplicity here we concentrate on the spherically symmetric case (distortion and rotation  can be included \cite{Ashtekar:2004gp} but this is not central to the argument presented here). 

Concretely, if one puts the characteristic data corresponding to the Schwarzschild horizon with area $A$ on $\Delta$ and then completes the characteristic data with suitable Schwarzschild data on $\delta$, then one recovers the Schwarzschild geometry as the unique solution of Einsteins equations (up to diffeomorphisms) in the domain of dependence of $\Delta\cup \delta$. This is a single point in the phase space  $\Gamma_A$ of spherically symmetric isolated horizons.
The infinite dimensional phase space $\Gamma_A$ is given by the set of solutions of Einstein's equations obtained by keeping the previous data fixed on $\Delta$ while setting arbitrary free characteristic data on $\delta$.
By fixing Schwarzschild data with (macroscopic) horizon area $A$ on $\Delta$ the definition introduces the coarse graining that is the cornerstone of the thermodynamical description. 

The phase space $\Gamma_A$ can be quantised according to the standard canonical prescription
once appropriate fundamental extended field variables are chosen \cite{Lewandowski:2005jk}. The Hilbert space 
$\sH$ of the system splits into two factors according to $\sH=\sH_{\Delta}\otimes \sH_{out}$, where $\sH_{\Delta}$ is the
a boundary Hilbert space (describing the degrees of freedom of the isolated horizon), and $\sH_{out}$ is the exterior volume Hilbert space (describing the physical degrees of freedom in the exterior of the BH horizon).
The latter is defined by solutions of the dynamical constraints in the bulk which are assumed to be expressible in terms of 
suitable combinations of spin network states (including matter field configurations supported on them). 
The Hilbert space $\sH_{\Delta}$ can be shown to be given by an $SU(2)$ Chern-Simons Hilbert space of a sphere $S^2$ with an arbitrary number of sources labelled by spins \cite{Engle:2009vc}\footnote{For a proposal of an intrinsic derivation of the BH physical states from the full theory see \cite{Sahlmann:2011xu}.}. Physical states satisfy a boundary constraint imposing the spins of the Chern-Simons sources to match the spins of the open links of bulk spin network states in $\sH_{out}$ \cite{ABK}.

The physical states described above are eigenstates of the IH horizon area  operator $\hat A_{\Delta}$ whose eigenvalues (being $\hat A_{\Delta}$ local) depend only on the spins labelling the Chern-Simons states (or equivalently the puncturing links of the bulk spinnetworks). More precisely, if $\{j_p\}$ denote the ensemble of spins at punctures labelled by an index $p$ then 
\be
\widehat A_{\Delta} \ |\psi^{\{j_p\}}_{out}  \rangle= 8\pi\gamma\ell_{Pl}^2\sum_{p}\sqrt{j_p(j_p+1)}\ |\psi^{\{j_p\}}_{out}  \rangle
\ee
where $|\psi^{\{j_p\}}_{out}   \rangle\in \sH_{\Delta}\otimes \sH_{out}$ and the subindex {\em out} schematically denotes all the quantum numbers corresponding to labels of a complete basis of local observables in the {\em outside}. Notice that these include the quantum numbers describing not only the space-time geometry but also the matter degrees of freedom.  The parameter $\gamma$ is the Immirzi parameter.
Entropy of the isolated Horizon is defined (in the microcanonical) approach as \be\label{mce}
S_{BH}=\log(\sN_{A})
\ee
where $\sN_A$ spin configurations satisfying the restriction. 
\begin{equation}
A-\delta\leq 8\pi\gamma\ell_{Pl}^2\sum_{p}\sqrt{j_p(j_p+1)}\leq A+\delta\,.
\label{QG_004}
\end{equation}
Notice that one is only counting horizon surface degrees of freedom that are 
compatible with the coarse graining condition defined by the macroscopic area $A$
that defines the isolated horizon.

\subsubsection*{Correlations across isolated horizons}

\begin{figure}[h]
\psfrag{A}{ \ {\em inside}}
\psfrag{B}{\,\! {\em outside}}\psfrag{C}{$\Delta$}
\includegraphics[width=5cm]{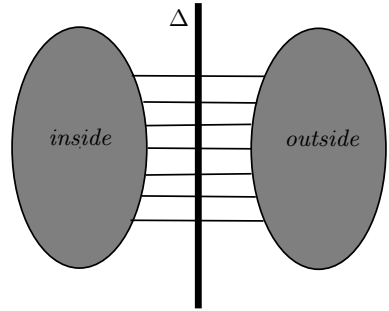}
\caption{The isolated horizon quantisation allows for correlations to be mediated only through the links of spin-networks 
crossing the horizon. This implies the equivalence between entanglement and statistical mechanical entropy.} \label{state}
\end{figure}

\begin{figure}[h]
\psfrag{mytag}{$\delta'$}
\psfrag{b}{$\Delta$}\psfrag{c}{$\delta$}
\includegraphics[width=7cm]{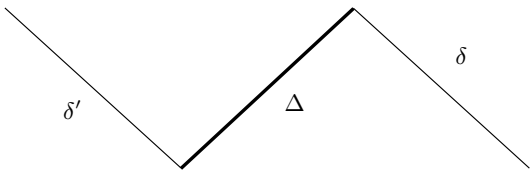}
\caption{A set of null surfaces with a non trivial domain of dependence containing the isolated horizon $\Delta$.} \label{two}
\end{figure}

A key point is that one could follow a similar procedure and construct the {\em inside} physical states.
Starting from the phase space $\Gamma'_A$ of the {\em inside} isolated horizon---defined now in terms of the same characteristic data on $\Delta$ and independent free data on $\delta'$ (see Figure \ref{two}).
One can thus define a Hilbert space $\sH'=\sH_{in}\otimes \sH_{\Delta}$ and, going through the very same 
prescription, obtain the physical states describing the fields on the {\em inside} of the isolated horizon.
Such states can be denoted $|\psi^{\{j_p\}}_{in}  \rangle\in \sH_{in}\otimes \sH_{\Delta}$ in analogy with the previous notation for {\em outside} physical states.
The horizon constraints imply that the number of punctures with their spins arriving at $\Delta$  from the {\em inside} match the number of punctures with their spin arriving at $\Delta$ from the {\em outside}. This implies that the physical states describing 
a neighbouring four dimensional region of an isolated horizon $\Delta$ can be graphically represented as in Figure \ref{state}, where the links connecting the {\em inside} with the {\em outside} are spin network edges labelled by spins that are not explicitly written for notational simplicity. 

At this stage it will be convenient to adopt a more compact notation 
\ba
\n |\psi^{\{j_p\}}_{in}  \rangle &\longrightarrow& |\psi^a_{in}\rangle \\
|\psi^{\{j_p\}}_{out}   \rangle  &\longrightarrow& |\psi^a_{out}\rangle 
\ea
where the single multy-index $a$ labels eigenvalues of the area and replaces the ensemble $\{j_p\}$. 
Then a state describing a four dimensional region of the space-time containing the horizon 
compatible with the coarse graining condition (\ref{QG_004}) can be written as follows 
\be \label{here}
|\Psi\rangle=\sum\limits_{a} \alpha_{ a } \,  |\psi^{a}_{in}\rangle\, |\psi^{a}_{out}\rangle, \ee 
where $|\psi^{a}_{in}\rangle$ and $|\psi^{a}_{out}\rangle$  denote physical states compatible with the boundary data $a$, and describing  interior and in the exterior state 
of matter and geometry of the BH respectively and the $\alpha_a$ are coefficients. 

The expression (\ref{here}) of the physical state follows directly from the definition of the isolated horizon quantization. The form of this equation implies that correlations  between the {\em outside} and the {\em inside} at Planckian scales are mediated by the spin-network links puncturing the separating boundary. The isolated horizon treatment restricts the vacuum correlations to be ultra-local at Planck scale\footnote{The heart of the point raised here might be related to properties of tensor network states and related notions where area law for entanglement entropy holds by construction. For a review and references see \cite{Eisert:2008ur}.}.

As described above, the bulk degrees of freedom in the isolated horizon framework
consist of entirely outgoing modes of gravitational and matter radiation. By bulk degrees of freedom here we mean those 
that are controlled by the Hamiltonian and diffeomorphism constraints on the interior of $\delta$.  These bulk degrees of freedom are ignored in the computation of BH entropy; such prescription is well justified physically by the discussion 
of the physics of gravitational collapse. However, notice that they are exactly of the same nature of those excitations leading to long range correlation contributions to the entanglement entropy in (\ref{dentada}). The separation of degrees of freedom necessary in the entanglement account of BH entropy and the one provided by the IH prescription are in this way compatible with each other.

\section{Entanglement entropy vs. statistical mechanical entropy}\label{sf}

Now we can compute the entanglement entropy associated to the local surface correlations across the horizon. We will do this first  in the case of a pure state of the form (\ref{here}). We will show that it is 
bounded by the micro canonical entropy (\ref{mce}); however, its precise value depends (as should be expected) on the values of the coefficients $\alpha_a$. As argued above the nature of the BH collapse is such that the $\alpha_a$ encode details of the microphysics that becomes unavailable in the coarse grained setting where the notion of BH entropy makes sense. Such ignorance of the microphysics leading to the macroscopic BH requires the introduction of an statistical mixture or mixed state. When this is done then the entanglement entropy 
of the physically appropriate mixed state is equal to the micro canonical entropy in LQG.

\subsubsection{Pure state entanglement entropy}

We start from (\ref{here}) and assume states to be normalized as follows: $\langle \psi_{out}^a|\psi_{out}^a\rangle=1$, $\langle \psi_{in}^a|\psi_{in}^a\rangle=1$, and  $\langle \Psi|\Psi\rangle=1$.
The pure state density matrix is
\be\label{pure}
\rho=|\Psi\rangle\langle\Psi|=\sum\limits_{a, a'} \alpha_a\bar\alpha_a'  \,  |\psi^{a}_{in}\rangle\, |\psi^{a}_{out}\rangle  \,  \langle \psi^{a'}_{in}|\, \langle \psi^{a'}_{out}|
\ee
The reduced density matrix, obtained by tracing over the {\em inside} degrees of freedom, becomes 
\ba
\n \rho_{out}&\equiv&\sum_{i,a} \langle a,i |\Psi\rangle\langle\Psi| a,i\rangle\\
 &=&\sum_{a,i} |\alpha_a|^2\beta_{ia}\bar\beta_{ia}\, |\psi^{a}_{out}\rangle\langle\psi^{a}_{out}| 
\ea
where $\beta_{ia}\equiv \langle a,i |\psi^a_{in}\rangle$. From this one gets \ba
\n \rho_{out}
&=&\sum\limits_a p_a |\psi^{a}_{out}\rangle\langle \psi^{a}_{out}|, 
\ea
with $p_a=|\alpha_a|^2$. The question now is what are the correct physical values of the probabilities $p_a$?
In the present context where we started from a pure state (\ref{pure}) the only thing we can say is that the (pure state) entanglement entropy \be S_{out}\equiv -{\rm Tr}[\rho_{out} \log(\rho_{out})]\le \log(\sN_A)=S_{BH},\ee 
i.e. it is bounded by the micro-canonical entropy (\ref{mce}). The equivalence can be achieved if  the physical states (pure states) describing a semiclassical black hole would have nearly to equal values of the probabilities $p_a=|\alpha_a|^2$ in the set of $a$ satisfying (\ref{QG_004}). 

The approximate equiprobability condition is certainly plausible considering that the recovering of semiclassical states compatible with the low energy continuum limit would require the contribution of the vast ensemble of eigenstates of the horizon area operator. Contributions might turn out to be approximately homogeneous in the range (\ref{QG_004}). Unfortunately, details on the way in which the continuum limit is to be recovered in the LQG framework are hard to quantify at present in the framework. However, we will argue in the following subsection that this perhaps not necessary.

\subsubsection{Mixed states entanglement entropy}

As we have argued in the introduction the coarse grained physics of the horizon is insensitive to the Planckian details that distinguish the different pure (micro) states  producing black holes with a definite final value of their extensive parameters. 
In such context the idea of reproducing {\em the} quantum state (as a pure state) reflecting the low energy properties of the suitable vacuum is unrealistic (as it would be to established the phase space coordinates of $10^{23}$ particles in a gas in equilibrium inside a box). As discussed before the only available information on the state producing the black hole is coarse grained (low resolution) information.
In such context the best one can do is to define a density matrix describing a mixed state corresponding to the statistical mixture of all the micro states---each of which would correspond to a distinguished initial pure state of matter and fields before collapse---satisfying our coarse graining criterion.

Consequently instead of (\ref{pure}) we take
\be
\rho=\sum\limits_{a, a'} p_{aa'}  \,  |\psi^{a}_{in}\rangle\, |\psi^{a}_{out}\rangle  \,  \langle \psi^{a'}_{in}|\, \langle \psi^{a'}_{out}|
\ee
as our starting point. Here $p_{aa'}$ is a positive definite matrix of coefficients describing the statistical mixture of states of the form (\ref{here}). The reduced density matrix becomes 
\ba
\n \rho_{out}&\equiv&\sum_{i,a} \langle a,i |\rho| a,i\rangle\\
&=&\sum_{a,i} p_{aa}\, |\psi^{a}_{out}\rangle\langle\psi^{a}_{out}| 
\ea
which, as a result of the partial trace, only depends on the diagonal elements $p_{aa}$.
The assumption of ergodicity (expected in a system with such huge number of microscopic degrees of freedom in equilibrium) together with
the idea that horizon area is the correct coarse graining observable naturally selects a $p_{aa}$ independent of the particular eigenvalue within the range
(\ref{QG_004}), hence $p_{aa}=\sN_A^{-1}$. From this is follows that
\be
S_{out}=S_{BH}.
\ee

Moreover, if a notion of local energy is available for those observers that are stationary with respect to the horizon one can 
find the equivalence with a canonical ensemble description.  
Assuming that the horizon is in thermal equilibrium with the {\em outside} for such observers, then the previous equiprobability criterion leads to the Boltzman probabilities $p_a=Z^{-1}\exp (-\beta E_a)$ where $\beta$ is inverse temperature and $E_a$ is the corresponding energy eigenvalue.  The Hamiltonian is the one generating a time flow
with respect to which the system is in local equilibrium \cite{Connes:1994hv}.
This is precisely the area Hamiltonian introduced in \cite{Frodden:2011eb}.
In such case we recover the canonical ensemble standard expression of the entropy.
The structure of physical states at the Planck  scale and the semiclassical input of the previous section imply that entanglement entropy is just the same as  statistical mechanical entropy in LQG.

\subsection*{Long range correlations}
Equation (\ref{here}) is central for establishing the relationship between statistical and entanglement entropy in the framework of loop quantum gravity. As argued, this is a consequence of the IH boundary condition (notice that the restriction imposed here on short range correlations are similar to those imposed in recent treatments \cite{Bianchi:2012ui}  or  \cite{Chirco:2014saa}, yet they are weaker in the sense that correlations in the UV are defined at a single horizon puncture).  The question that remains is wether such condition is compatible with the form of  {\em inside-outside} correlations expected in semiclassical states describing the near horizon geometry. 

Even thought it would be necessary to leave the framework of
quantum isolated horizons in order to describe situations where the bulk physics is relevant and {\em in-out} exchanges are allowed, it seems reasonable to expect that the dominant contributions to BH entropy (coming from the UV correlations close to the horizon) should be correctly captured by the present treatment. As argued in Section \ref{next}, long range correlations contribute to changes in entanglement entropy that can correctly be accounted for by the notion of relative entropy which is insensitive to the UV structure of spacetime. 


The description of BH entropy from the entanglement perspective necessarily requires a separation of scales where UV correlations close to the horizon dominate over IR long range correlations. Such separation of scales is natural from the classical physics of gravitational collapse, and the definition of the coarse graining that associates micro states to Planckian details of the horizon physics. Quantum isolated horizons are designed to model these fundamental excitations of the horizon degrees of freedom. They provide at the same time the natural separation of scales and the UV structure of geometry near the horizon.

%
%
%
%
\section{Discussion}\label{discu}
 
We have  illustrated how, by generalising the discussion of \cite{Bianchi:2012br} to static black hole backgrounds,  without the appropriate separation of degrees of freedom, entanglement entropy contains bulk terms which in perturbation theory represent leakage of entropy at infinity. Even when these flow terms can be interpreted thermodynamically in a suitable sense (see eq. (\ref{huno}) and footnote \ref{foot}) they are not associated to BH entropy. This is not surprising as it is well know that entanglement across the horizon has bulk contributions which strongly depend on the nature of the quantum state considered due to long range correlations which are unrestricted in QFT. For the entanglement approach to black hole entropy is essential to have the means to separate the wheat of local UV contributions across the BH horizon from the straw of long range correlations. 

Another important aspect is that relative entropy (for perturbations of the vacuum state) is by definition insensitive to the UV microphysics that we expect to be at the root of a fundamental account for BH entropy. 
A counterpart of this is the fact that changes of entanglement entropy and changes in the horizon law are related by the use of the semiclassical Einsteins equations where geometry remains classical (low energy physics governs the relationship (\ref{dent})). The fact that the inclusion of the (semiclassical) gravitational coupling eliminates the species problem in a thermodynamical sense. This is an encouraging indication which needs to be realized  at the fundamental level (statistically) by the description of the would-be-divergent UV contributions of matter and its background independent coupling to geometry in quantum gravity.   

The framework of quantum isolated horizon accounts for the UV excitations of the horizon geometric and non geometric degrees of freedom (including matter). These are the micro states in an ensemble defined by the macroscopic parameters associated to a stationary semiclassical horizon. 
Once the appropriate physical UV degrees of freedom are identified, entanglement entropy and statistical mechanical entropy are the same. Conversely, the physical states selected by the isolated horizon boundary condition present {\em inside-outside} correlations leading  to the maximum entanglement of the geometric degrees of freedom (under the standard assumption of equal probability distribution for states satisfying (\ref{QG_004})). In such situation entanglement entropy coincides with the logarithm of the number of micro states compatible with the coarse graining macroscopic parameters that define the statistical ensemble. 

The present discussion naturally leads to a possibly 
clarifying perspective on the information loss discussion. In Section \ref{info} we described this perspective 
based on the possibility of having purifying correlations between the low energy degrees of freedom of continuous fields and the UV discrete gravitational degrees of freedom that are hidden to standard low energy semiclassical observers. Such possibility is allowed by quantum gravity theories where continuum physics is only recovered under appropriate coarse graining of local observables.
This view is complementary to the one expressed in \cite{Ashtekar:2005cj} where the fate of classical singularities in quantum gravity is a key element of the discussion.

\section{Acknowledgements}

I am grateful for the interactions with I. Agullo, S. Ariwahjoedi, F. Barbero, G. Chirco, H. Haggard,  A. Riello, 
C. Rovelli, J. Pullin and D. Sudarsky. This work has been carried out thanks to the support of the OCEVU Labex (ANR-11-LABX-0060) and the A*MIDEX project (ANR-11-IDEX-0001-02) funded by the "Investissements d'Avenir" French government program managed by the ANR.



\begin{thebibliography}{10}

\bibitem{Perez:2004hj}
Alejandro Perez.
\newblock {Introduction to loop quantum gravity and spin foams}.
\newblock 2004.

\bibitem{lqg1}
Carlo Rovelli.
\newblock {Quantum gravity}.
\newblock {\em Cambridge University Press (2004)}.

\bibitem{lqg2}
Thomas Thiemann.
\newblock {Modern canonical quantum general relativity}.
\newblock 2001.

\bibitem{Rovelli:1996dv}
Carlo Rovelli.
\newblock {Black hole entropy from loop quantum gravity}.
\newblock {\em Phys.Rev.Lett.}, 77:3288--3291, 1996.

\bibitem{ABK}
Abhay Ashtekar, John Baez, and Kirill Krasnov.
\newblock {Quantum Geometry of Isolated Horizons and Black Hole Entropy}.
\newblock {\em Adv.Theor.Math.Phys.}, 4:1--94, 2000.

\bibitem{Meissner:2004ju}
Krzysztof~A. Meissner.
\newblock Black-hole entropy in loop quantum gravity.
\newblock {\em Class. Quant. Grav.}, 21:5245–--5251, 2004.

\bibitem{Agullo:2008yv}
Iv\'an Agull\'o, J.~Fernando Barbero~G., Jacobo D\'{\i}az-Polo, Enrique
  F.~Borja, and Eduardo J.~S. Villase\~nor.
\newblock Black hole state counting in loop quantum gravity: A
  number-theoretical approach.
\newblock {\em Phys. Rev. Lett.}, 100:211301--1--4, 2008.

\bibitem{Date:2008rb}
Ghanashyam Date, Romesh~K. Kaul, and Sandipan Sengupta.
\newblock {Topological Interpretation of Barbero-Immirzi Parameter}.
\newblock {\em Phys.Rev.}, D79:044008, 2009.

\bibitem{Rezende:2009sv}
Danilo~Jimenez Rezende and Alejandro Perez.
\newblock {4d Lorentzian Holst action with topological terms}.
\newblock {\em Phys.Rev.}, D79:064026, 2009.

\bibitem{Jacobson:2007uj}
Ted Jacobson.
\newblock {Renormalization and black hole entropy in Loop Quantum Gravity}.
\newblock {\em Class.Quant.Grav.}, 24:4875--4879, 2007.

\bibitem{Ghosh:2013iwa}
Amit Ghosh, Karim Noui, and Alejandro Perez.
\newblock {Statistics, holography, and black hole entropy in loop quantum
  gravity}.
\newblock 2013.

\bibitem{Agullo:2012fc}
Ivan Agullo, Abhay Ashtekar, and William Nelson.
\newblock {Extension of the quantum theory of cosmological perturbations to the
  Planck era}.
\newblock {\em Phys.Rev.}, D87(4):043507, 2013.

\bibitem{Bombelli:1986rw}
Luca Bombelli, Rabinder~K. Koul, Joohan Lee, and Rafael~D. Sorkin.
\newblock {A Quantum Source of Entropy for Black Holes}.
\newblock {\em Phys.Rev.}, D34:373--383, 1986.

\bibitem{Frodden:2012dq}
Ernesto Frodden, Marc Geiller, Karim Noui, and Alejandro Perez.
\newblock {Black Hole Entropy from complex Ashtekar variables}.
\newblock 2012.

\bibitem{Han:2014xna}
Muxin Han.
\newblock {Black Hole Entropy in Loop Quantum Gravity, Analytic Continuation,
  and Dual Holography}.
\newblock 2014.

\bibitem{Ghosh:2014rra}
Amit Ghosh and Daniele Pranzetti.
\newblock {CFT/Gravity Correspondence on the Isolated Horizon}.
\newblock 2014.

\bibitem{Pranzetti:2013lma}
Daniele Pranzetti.
\newblock {Black hole entropy from KMS-states of quantum isolated horizons}.
\newblock {\em Phys.Rev.}, D89:104046, 2014.

\bibitem{Bianchi:2012br}
Eugenio Bianchi.
\newblock {Horizon entanglement entropy and universality of the graviton
  coupling}.
\newblock 2012.

\bibitem{Bodendorfer:2014fua}
Norbert Bodendorfer.
\newblock {A note on entanglement entropy and quantum geometry}.
\newblock 2014.

\bibitem{Husain:1998zw} 
  V.~Husain,
  \newblock{Apparent horizons, black hole entropy and loop quantum gravity}
  Phys.\ Rev.\ D {\bf 59}, 084019 (1999)
  [gr-qc/9806115].

\bibitem{Donnelly:2008vx}
William Donnelly.
\newblock {Entanglement entropy in loop quantum gravity}.
\newblock {\em Phys.Rev.}, D77:104006, 2008.

\bibitem{Wald:1993nt}
Robert~M. Wald.
\newblock {Black hole entropy is the Noether charge}.
\newblock {\em Phys.Rev.}, D48:3427--3431, 1993.

\bibitem{Solodukhin:2011gn}
Sergey~N. Solodukhin.
\newblock {Entanglement entropy of black holes}.
\newblock {\em Living Rev.Rel.}, 14:8, 2011.

\bibitem{Wald:1995yp}
Robert~M. Wald.
\newblock {Quantum field theory in curved space-time and black hole
  thermodynamics}.
\newblock 1995.

\bibitem{Casini:2008cr}
H.~Casini.
\newblock {Relative entropy and the Bekenstein bound}.
\newblock {\em Class.Quant.Grav.}, 25:205021, 2008.

\bibitem{Frodden:2011eb}
Ernesto Frodden, Amit Ghosh, and Alejandro Perez.
\newblock {Quasilocal first law for black hole thermodynamics}.
\newblock {\em Phys.Rev.}, D87:121503, 2013.

\bibitem{Kay:1988mu}
Bernard~S. Kay and Robert~M. Wald.
\newblock {Theorems on the Uniqueness and Thermal Properties of Stationary,
  Nonsingular, Quasifree States on Space-Times with a Bifurcate Killing
  Horizon}.
\newblock {\em Phys.Rept.}, 207:49--136, 1991.

\bibitem{huang}
Zichang Huang, Alejandro Perez, and Simone Speziale.
\newblock {Hawking radiaiton: an exactly solvable model}.
\newblock {\em in progress}.

\bibitem{Wald:1999vt}
Robert~M. Wald.
\newblock {The thermodynamics of black holes}.
\newblock {\em Living Rev.Rel.}, 4:6, 2001.

\bibitem{JACOBSON:2013ewa}
Ted Jacobson.
\newblock {Boundary unitarity and the black hole information paradox}.
\newblock {\em Int.J.Mod.Phys.}, D22:1342002, 2013.

\bibitem{Varadarajan:1999it}
Madhavan Varadarajan.
\newblock {Fock representations from U(1) holonomy algebras}.
\newblock {\em Phys.Rev.}, D61:104001, 2000.

\bibitem{Ashtekar:2001xp}
Abhay Ashtekar and Jerzy Lewandowski.
\newblock {Relation between polymer and Fock excitations}.
\newblock {\em Class.Quant.Grav.}, 18:L117--L128, 2001.

\bibitem{Ashtekar:2002sn}
Abhay Ashtekar, Stephen Fairhurst, and Joshua~L. Willis.
\newblock {Quantum gravity, shadow states, and quantum mechanics}.
\newblock {\em Class.Quant.Grav.}, 20:1031--1062, 2003.

\bibitem{Eisert:2008ur}
J.~Eisert, M.~Cramer, and M.B. Plenio.
\newblock {Area laws for the entanglement entropy - a review}.
\newblock {\em Rev.Mod.Phys.}, 82:277--306, 2010.

\bibitem{Bousso:2014sda}
Raphael Bousso, Horacio Casini, Zachary Fisher, and Juan Maldacena.
\newblock {Proof of a Quantum Bousso Bound}.
\newblock 2014.

\bibitem{Agullo:2009wt}
Ivan Agullo, Jose Navarro-Salas, Gonzalo~J. Olmo, and Leonard Parker.
\newblock {Insensitivity of Hawking radiation to an invariant Planck-scale
  cutoff}.
\newblock {\em Phys.Rev.}, D80:047503, 2009.

\bibitem{Gambini:2013nea}
Rodolfo Gambini and Jorge Pullin.
\newblock {Hawking radiation from a spherical loop quantum gravity black hole}.
\newblock 2013.

\bibitem{Collins:2004bp}
John Collins, Alejandro Perez, Daniel Sudarsky, Luis Urrutia, and Hector
  Vucetich.
\newblock {Lorentz invariance and quantum gravity: an additional fine-tuning
  problem?}
\newblock {\em Phys.Rev.Lett.}, 93:191301, 2004.

\bibitem{Polchinski:2011za}
Joseph Polchinski.
\newblock {Comment on [arXiv:1106.1417] 'Small Lorentz violations in quantum
  gravity: do they lead to unacceptably large effects?'}.
\newblock {\em Class.Quant.Grav.}, 29:088001, 2012.

\bibitem{AKLR}
Abhay Ashtekar and Badri Krishnan.
\newblock Isolated and dynamical horizons and their applications.
\newblock {\em Living Rev. Relativity}, 7, 2004.

\bibitem{Ashtekar:2004gp}
Abhay Ashtekar, Jonathan Engle, Tomasz Pawlowski, and Chris Van Den~Broeck.
\newblock {Multipole moments of isolated horizons}.
\newblock {\em Class.Quant.Grav.}, 21:2549--2570, 2004.

\bibitem{Lewandowski:2005jk}
Jerzy Lewandowski, Andrzej Okolow, Hanno Sahlmann, and Thomas Thiemann.
\newblock {Uniqueness of diffeomorphism invariant states on holonomy-flux
  algebras}.
\newblock {\em Commun.Math.Phys.}, 267:703--733, 2006.

\bibitem{Engle:2009vc}
Jonathan Engle, Karim Noui, and Alejandro Perez.
\newblock Black hole entropy and {SU}(2) {C}hern-{S}imons theory.
\newblock {\em Phys. Rev. Lett.}, 105, 2010.

\bibitem{Sahlmann:2011xu}
Hanno Sahlmann.
\newblock {Black hole horizons from within loop quantum gravity}.
\newblock {\em Phys.Rev.}, D84:044049, 2011.

\bibitem{Connes:1994hv}
A.~Connes and Carlo Rovelli.
\newblock {Von Neumann algebra automorphisms and time thermodynamics relation
  in general covariant quantum theories}.
\newblock {\em Class.Quant.Grav.}, 11:2899--2918, 1994.

\bibitem{Bianchi:2012ui}
Eugenio Bianchi.
\newblock {Entropy of Non-Extremal Black Holes from Loop Gravity}.
\newblock 2012.

\bibitem{Chirco:2014saa}
Goffredo Chirco, Hal~M. Haggard, Aldo Riello, and Carlo Rovelli.
\newblock {Spacetime thermodynamics without hidden degrees of freedom}.
\newblock 2014.

\bibitem{Ashtekar:2005cj}
Abhay Ashtekar and Martin Bojowald.
\newblock {Black hole evaporation: A Paradigm}.
\newblock {\em Class.Quant.Grav.}, 22:3349--3362, 2005.

\end{thebibliography}
%
%
%
%

\end{document}